\documentclass{article}

\usepackage{epsfig,float}
\newcommand{\bfr}{\begin{flushright}}
\newcommand{\efr}{\end{flushright}}
 
\begin{document}
\title{Timelike vs spacelike DVCS from JLab, Compass to  ultraperipheral collisions and AFTER@LHC
\thanks{Presented at  the Low x workshop, May 30 - June 4 2013, Rehovot and
Eilat, Israel}
}
\author{H. Moutarde $^1$,
B. Pire $^2$, F. Sabati\'e $^1$, L. Szymanowski $^3$ and J. Wagner $^3$\\
{\small
1. Irfu-SPhN,  CEA, Saclay, France}\\
{\small 2. CPhT, \'Ecole polytechnique, CNRS, 91128 Palaiseau, France}\\
{\small 3. National Centre for Nuclear Research (NCBJ), Warsaw, Poland}\\
\smallskip\\
}
\date{\today
}
\maketitle
\begin{abstract}
Timelike and spacelike virtual Compton scattering  in the generalized Bjorken scaling regime are complementary tools to access generalized parton distributions. We stress that the gluonic contributions are by no means negligible, even in the medium energy range which will be studied intensely at JLab12 and in the COMPASS-II experiment at CERN. Ultraperipheral collisions with proton or ion beams may also be used at RHIC and at collider or fixed target experiments at LHC.
\\

\end{abstract}

\section{Introduction}

In the  collinear factorization framework the scattering amplitudes for  exclusive processes such as deeply virtual Compton scattering (DVCS) \cite{historyofDVCS} 
\begin{equation}
\gamma^*(q_{in}) N(P) \to \gamma(q_{out}) N'(P'=P+\Delta) \,,~~~~~q_{in}^2 =-Q^2< 0,~~~~~q_{out}^2 =0\,,
\end{equation}
and  its crossed reaction, timelike Compton scattering (TCS) \cite{Berger:2001xd}
\begin{equation}
\gamma(q_{in}) N(P)\to \gamma^*(q_{out}) N'(P'=P+\Delta)\,,~~~~~q_{in}^2 =0,~~~~~q_{out}^2 =Q^2>0 \,,
\end{equation}
have been shown to factorize in specific kinematical regions, provided a large scale controls the separation of short distance dominated partonic subprocesses and long distance hadronic matrix elements, the generalized 
quark and gluon distributions (GPDs)  \cite{review} which encode much information on the partonic content of nucleons.
After factorization, the DVCS (and similarly TCS) amplitudes are written in terms of  Compton form factors  (CFF) $\mathcal{H}$, $\mathcal{E}$ and $\widetilde {\mathcal{H}}$, $\widetilde {\mathcal{E}}$ , as :
\begin{eqnarray}
\mathcal{A}^{\mu\nu}(\xi,t) &=& - e^2 \frac{1}{(P+P')^+}\, \bar{u}(P^{\prime}) 
\Big[\,
   g_T^{\mu\nu} \, \Big(
      {\mathcal{H}(\xi,t)} \, \gamma^+ +
      {\mathcal{E}(\xi,t)} \, \frac{i \sigma^{+\rho}\Delta_{\rho}}{2 M}
   \Big) \\ &&\phantom{AAAAAAAAaaAA}
   +i\epsilon_T^{\mu\nu}\, \Big(
    {\widetilde{\mathcal{H}}(\xi,t)} \, \gamma^+\gamma_5 +
      {\widetilde{\mathcal{E}}(\xi,t)} \, \frac{\Delta^{+}\gamma_5}{2 M}
    \Big)
\,\Big] u(P)  \nonumber \, ,
\label{eq:amplCFF}
\end{eqnarray}
with the CFFs 
defined  as
\begin{eqnarray}
\mathcal{H}(\xi,t) &=& + \int_{-1}^1 dx \,
\left(\sum_q T^q(x,\xi)H^q(x,\xi,t)
 + T^g(x,\xi)H^g(x,\xi,t)\right) \; , \nonumber \\
\widetilde {\mathcal{H}}(\xi,t) &=& - \int_{-1}^1 dx \,
\left(\sum_q \widetilde {T}^q(x,\xi)\widetilde {H}^q(x,\xi,t) 
+\widetilde {T}^g(x,\xi)\widetilde {H}^g(x,\xi,t)\right).
\label{eq:CFF}
\end{eqnarray}
We report in Sect. 2 on a recent  NLO analysis \cite{PSW2, Moutarde:2013qs} of DVCS and TCS amplitudes., and make a few remarks on the study of TCS in ultraperipheral collisions at hadron colliders (Sect. 3) and at fixed target experiments at LHC (Sect. 4).


\begin{figure}[h]
\begin{center}
\includegraphics[width= 0.4\textwidth]{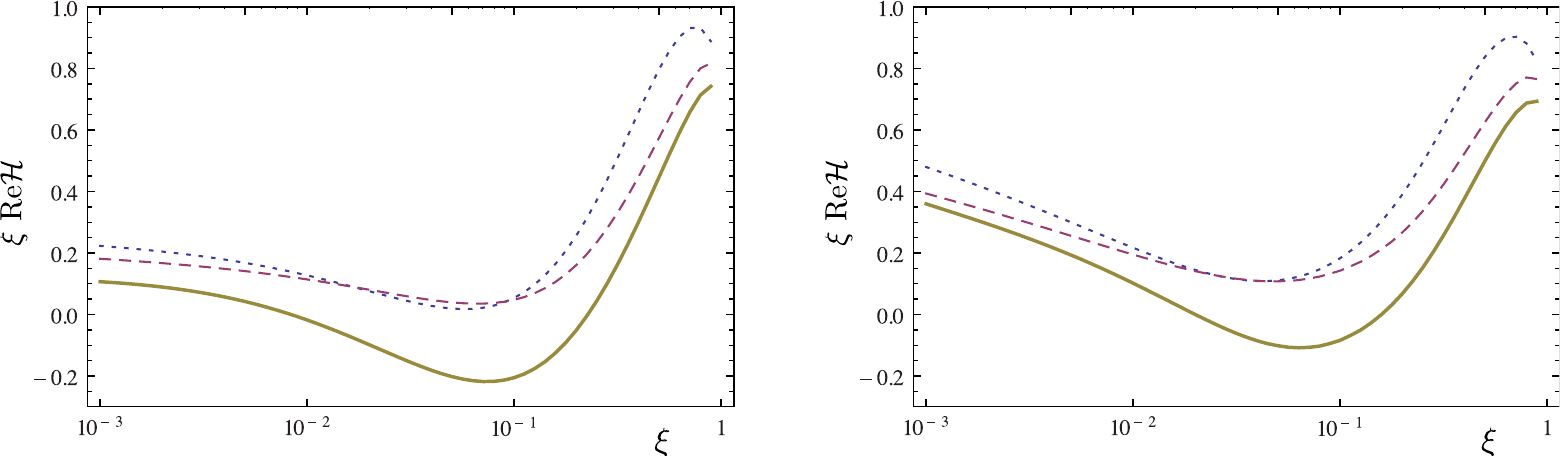} ~~~~~~  \includegraphics[width= 0.4\textwidth]{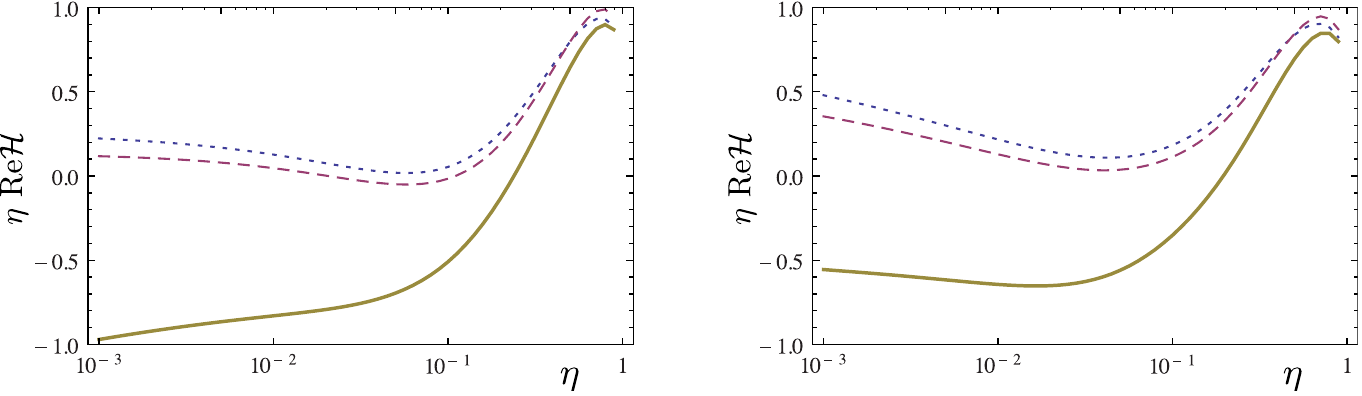} 
  \includegraphics[width= 0.4\textwidth]{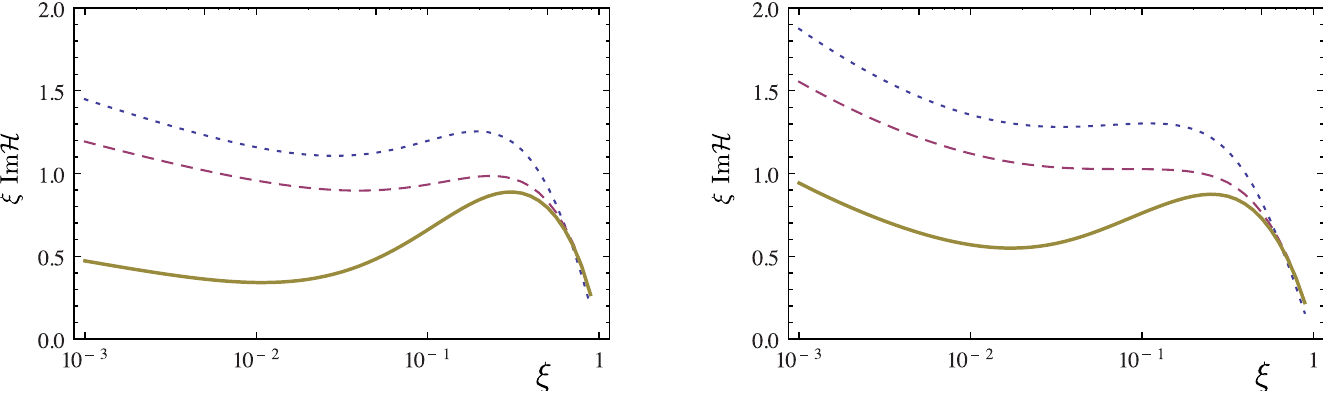}  ~~~~~~  \includegraphics[width= 0.4\textwidth]{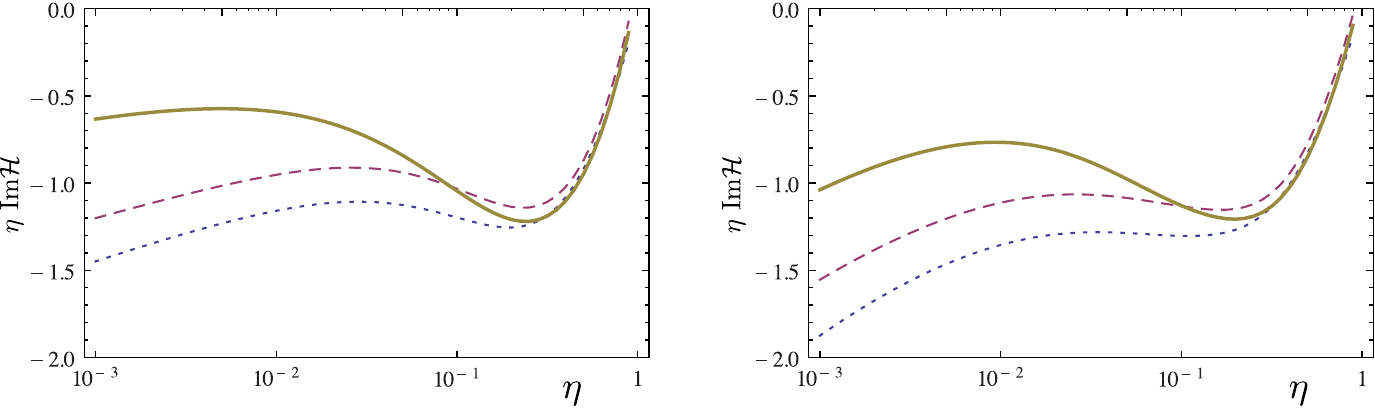} 
\caption{The real (upper panels) and imaginary (lower pannels) parts of the {\it spacelike} $\mathcal{H}(\xi)$ (1$^{st}$ and 2$^{nd}$ columns) and {\it timelike} $\mathcal{H}(\eta)$ (3$^{rd}$ and 4$^{th}$ columns) Compton Form Factor multiplied by $\xi$ (or $\eta$),  in the double distribution model based on GK (1$^{st}$ and 3$^{rd}$ columns) and MSTW08 (2$^{nd}$ and 4$^{th}$ columns) parametrizations, for $
\mu_F^2=Q^2=4 GeV^2$ and $t= -0.1 GeV^2$. In all plots, the LO result is shown as the dotted line, the full NLO result by the solid line and the NLO result without the gluonic contribution as the dashed line.}
\label{fig:DVCSRe2x2}
\end{center}
\end{figure}

\section{On the importance of gluonic contributions
\label{sec2}}
\subsection{Gluonic effects to Compton form factors} 
\begin{figure}[t]
\begin{center}
 \includegraphics[width=7cm]{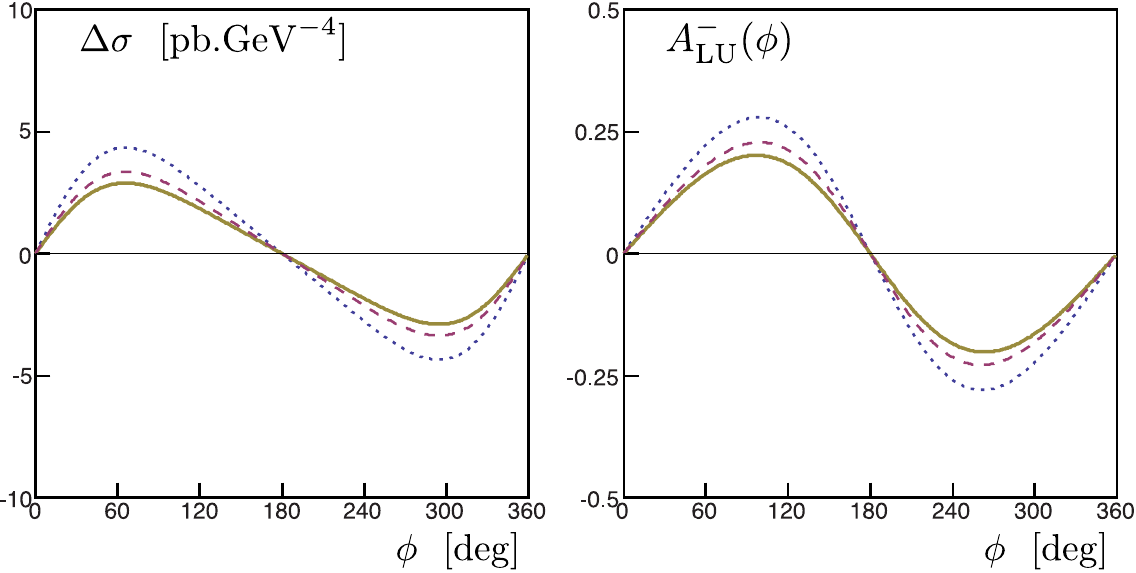} 
\caption{The difference of DVCS cross sections for opposite lepton helicities in pb/GeV$^4$ (left) and the corresponding asymmetry (right),  as a function of $\phi $    for $E_e=11 GeV, \mu_F^2=Q^2=4$~GeV$^2$ and $t= -0.2$~GeV$^2$. The GPD $H(x,\xi,t)$ is parametrized by the GK model. The contributions from other GPDs are not included.  }
\label{fig:c1}
\end{center}
\end{figure}
TCS and DVCS amplitudes are identical (up to a complex conjugation) at lowest order in $\alpha_S$ but differ at next to leading order, in particular because of the quite different analytic structure of the scattering amplitudes of these reactions. Indeed, the production of a timelike photon enables the production of intermediate states in some channels which are kinematically forbidden in the DVCS case. This opens the way to new absorptive parts of the amplitude.  To estimate Compton Form Factors (CFF), we use  the  NLO calculations of the coefficient functions which have been calculated in the DVCS case in the early days of GPD studies and more recently for  the TCS case \cite{PSW2}, the two results being simply related thanks to the analyticity (in $Q^2$) properties of the amplitude \cite{MPSW}:
\begin{eqnarray}
^{TCS}T(x,\eta) = \pm \left(^{DVCS}T(x,\xi=\eta) +  i \pi C_{coll}(x,\xi = \eta)\right)^* \,,
\label{eq:TCSvsDVCS}
\end{eqnarray}
where $+$~$(-)$ sign corresponds to vector (axial) case.

Using two GPD models based on Double Distributions (DDs), as discussed in detail in Ref.  \cite{Moutarde:2013qs} : the Goloskokov-Kroll (GK) model \cite{GK3} and a model based on the MSTW08 PDF parametrization \cite{Martin:2009iq}, we get the results shown in Fig. 1 for the real and imaginary parts of the spacelike and timelike dominant CFF $\mathcal{H}(\xi,t) $ and $\mathcal{H}(\eta,t) $.  Comparing dashed and solid lines in the upper panels, one sees that gluonic contributions are so important that they even change the sign of the real part of the CFF, and are dominant for almost all values of the skewness parameter. A milder conclusion arises from a similar comparison of the lower panels; the gluonic contribution to  the imaginary part of the CFF remains sizeable for values of the skewness parameter up to $0.3$.
\begin{figure}[h]
\begin{center}
  \includegraphics[width=10 cm]{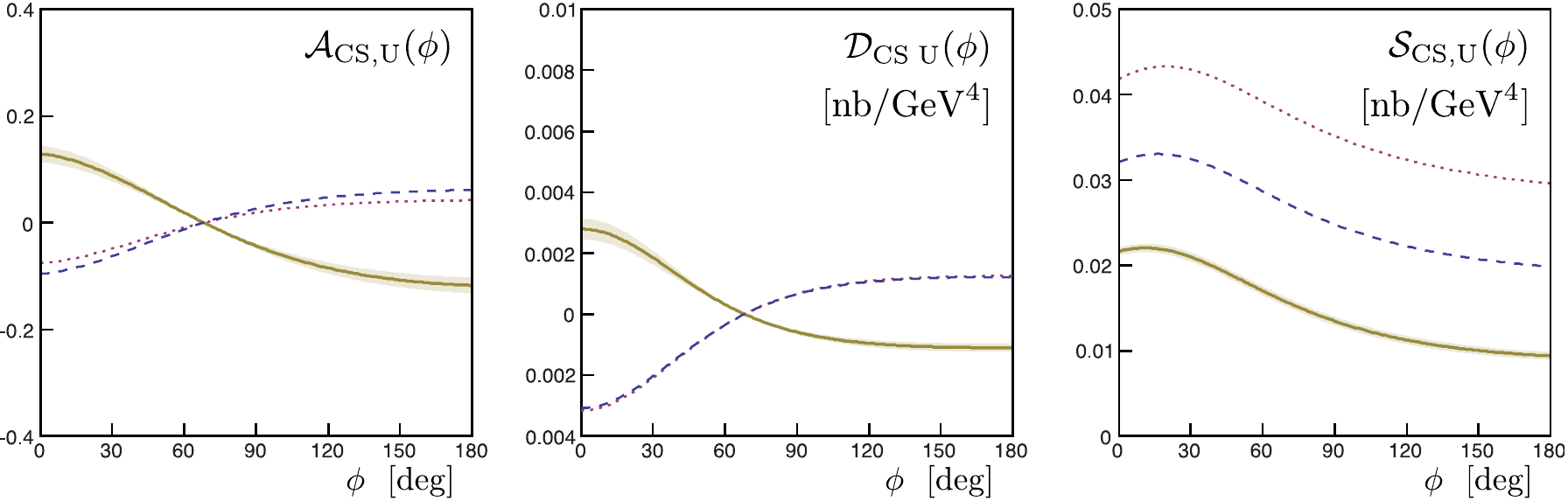}  
\caption{The DVCS observables for the COMPASS experiment, from left to right, mixed charge-spin asymmetry, mixed charge-spin difference and mixed charge-spin sum (in nb/GeV$^4$). The kinematical point is chosen as $ \xi=0.05, Q^2=4$~GeV$^2$, $t=-0.2$~GeV$^2$.  The GPD $H(x,\xi,t)$ is parametrized in the double distribution model based on the MSTW08 parametrization. The contributions from other GPDs are not included.}
\label{fig:compass}
\end{center}
\end{figure}

\subsection{Gluonic effects to  DVCS  observables}
The effects of NLO contributions to some of the DVCS observables at moderate energies are exemplified in Fig. \ref{fig:c1} and Fig. \ref{fig:compass} which show specific   observables to be measured at JLab and COMPASS. The difference between the dotted and solid lines demonstrates that NLO contributions are important, whereas the difference between the dashed and solid lines shows that gluon contributions should not be forgotten even at low energy when a precise data set is analyzed.

\section{TCS in ultraperipheral collisions 
\label{sec3}}
Timelike Compton scattering in ultraperipheral collisions at hadron 
colliders opens a new way to measure generalized parton distributions, in particular for very small values of the skewness parameter.

We estimated \cite{PSW} the different contributions to the lepton pair cross section for 
ultraperipheral collisions at the LHC.  Since the
cross sections decrease rapidly with $Q^2$, we are interested in the kinematics of moderate $Q^2$, 
say a few GeV$^2$, and large energy, thus very small values of $\eta$. 
Note however that for a given proton energy the photon flux is higher at 
smaller photon energy.

\noindent
 The Bethe-Heitler amplitude grows much when small $\theta$ angles are allowed.
In the following we will use the limits $[\pi/4,3 \pi/4]$ where the Bethe-Heitler cross section is sufficiently big but does not dominate too much over the Compton process. The Bethe-Heitler cross section integrated over $\theta \in [\pi/4,3
\pi/4]$, $\phi \in [0, 2\pi]$ , $Q^2 \in [4.5,5.5]GeV^2$, $|t| \in [0.05,0.25] GeV^2$, as a function of $\gamma p$  energy squared $s$ is  in the limit of large
 $s$  constant and equals $28.4 $ pb. 

\noindent
Since the amplitudes for the Compton and Bethe-Heitler
processes transform with opposite signs under reversal of the lepton
charge,  the interference term between TCS and BH is
odd under exchange of the $\ell^+$ and $\ell^-$ momenta.
It is thus possible to  project out the interference term through a clever use of
 the angular distribution of the lepton pair. 
The interference part of the cross-section for $\gamma p\to \ell^+\ell^-\, p$ with 
unpolarized protons and photons has a characteristic ($\theta, \phi$) dependence given  by (see details in \cite{PSW})
\begin{eqnarray}
   \label{intres}
\frac{d \sigma_{INT}}{dQ^2\, dt\, dcos\theta\, d\phi}
= {}-
\frac{\alpha^3_{em}}{4\pi s^2}\, \frac{1}{-t}\, \frac{M}{Q}\,
\frac{1}{\tau \sqrt{1-\tau}}\,
  cos\phi \frac{1+cos^2\theta}{sin\theta}
     {\cal R}e {\cal M} \; ,\nonumber
\end{eqnarray}
with 
\begin{equation}
\label{mmimi}
{\cal M} = \frac{2\sqrt{t_0-t}}{M}\, \frac{1-\eta}{1+\eta}\,
\left[ F_1 {\cal H}_1 - \eta (F_1+F_2)\, \tilde{\cal H}_1 -
\frac{t}{4M^2} \, F_2\, {\cal E}_1 \,\right],
\nonumber
\end{equation}
where $\tau=Q^2/s $, $-t_0 = 4\eta^2 M^2 /(1-\eta^2)$ and ${\cal H},  \tilde{\cal H}, {\cal E}$ are Compton form factors.
 With the integration limits symmetric about $\theta=\pi/2$ the interference
term changes sign under $\phi\to \pi+\phi$ due to charge conjugation,
whereas the TCS and BH cross sections do not. One may thus extract the 
Compton amplitude through a study of
$\int\limits_0^{2\pi}d\phi\,cos \phi \frac{d\sigma}{d\phi}$. 


\begin{figure*}
  \includegraphics[width=0.5\textwidth]{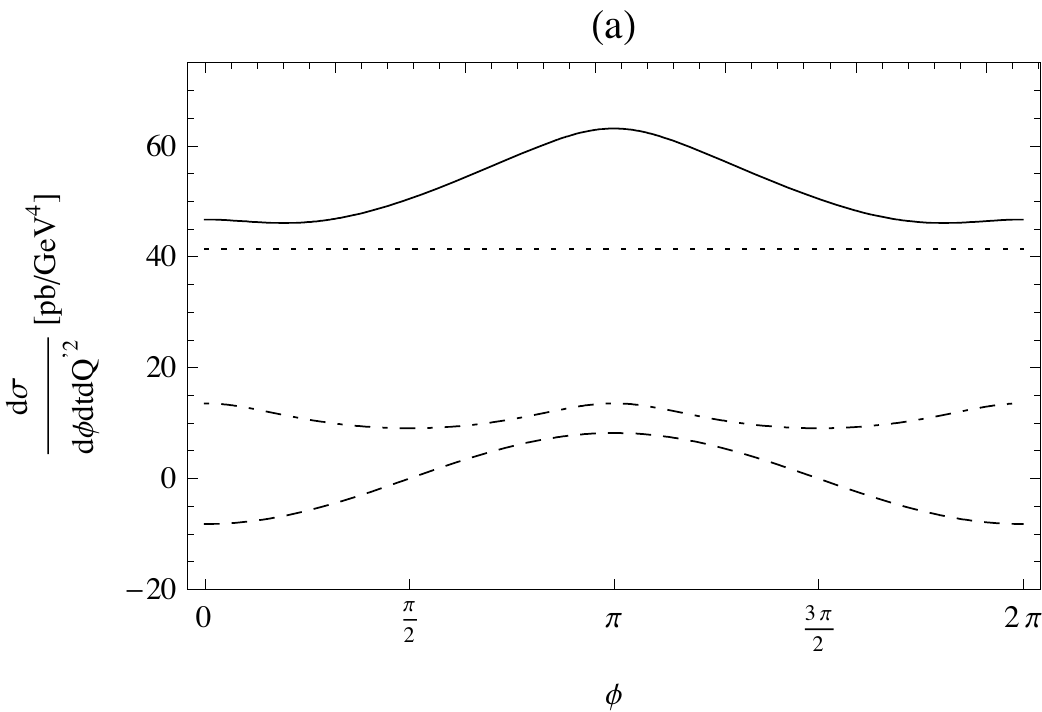}
\includegraphics[width=0.5\textwidth]{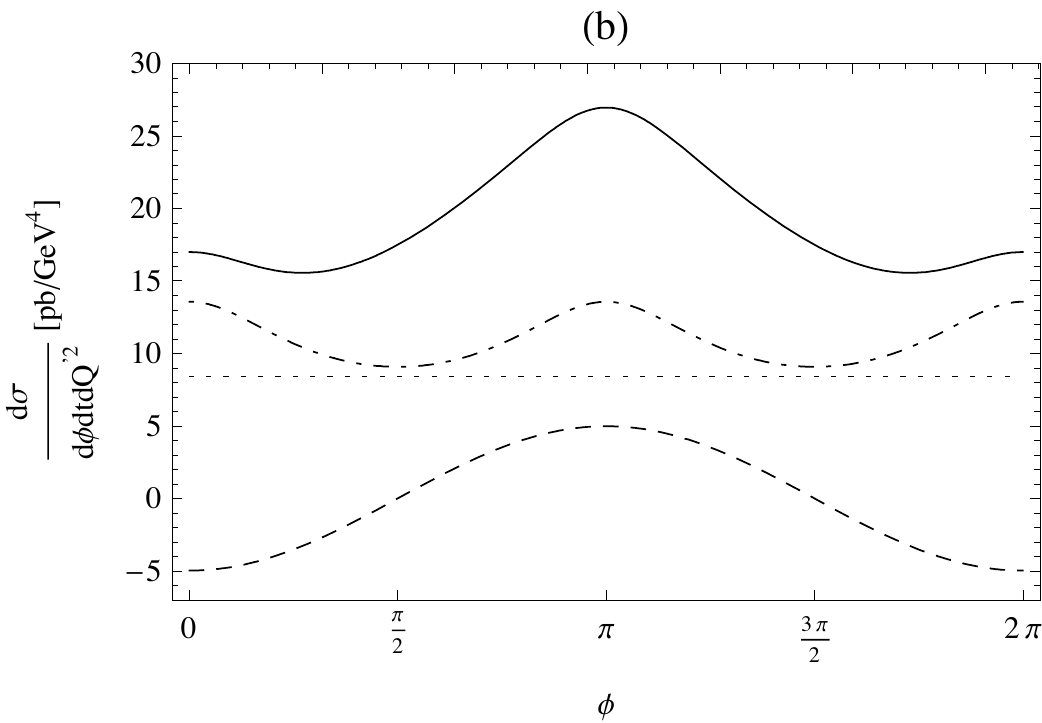} 
\caption{The lepton pair production  differential cross sections (solid lines) for $t =-0.2GeV^2$, ${Q}^2 =5 GeV^2$  integrated 
over $\theta = [\pi/4,3\pi/4]$, as a function of $\phi$, for $s=10^7 GeV^2$ (a), 
$s=10^5GeV^2$(b)
 with $\mu_F^2 = 5 GeV^2$. We also display  the
Compton (dotted), Bethe-Heitler (dash-dotted) and Interference (dashed) contributions.}
\label{Interf}
\end{figure*}

In Fig. \ref{Interf} we show the interference contribution to the cross section in comparison to the Bethe-Heitler and Compton processes, for various values of photon proton  energy 
squared $s = 10^7 GeV^2,10^5 GeV^2$. We observe that for 
larger energies the  
Compton process dominates, whereas for $s=10^5 GeV^2$ all contributions are comparable. 

\section{Ultraperipheral collisions in a high-energy fixed-target experiment AFTER@LHC}

\begin{figure}[h]
\begin{center}
   \includegraphics[width= 0.45\textwidth]{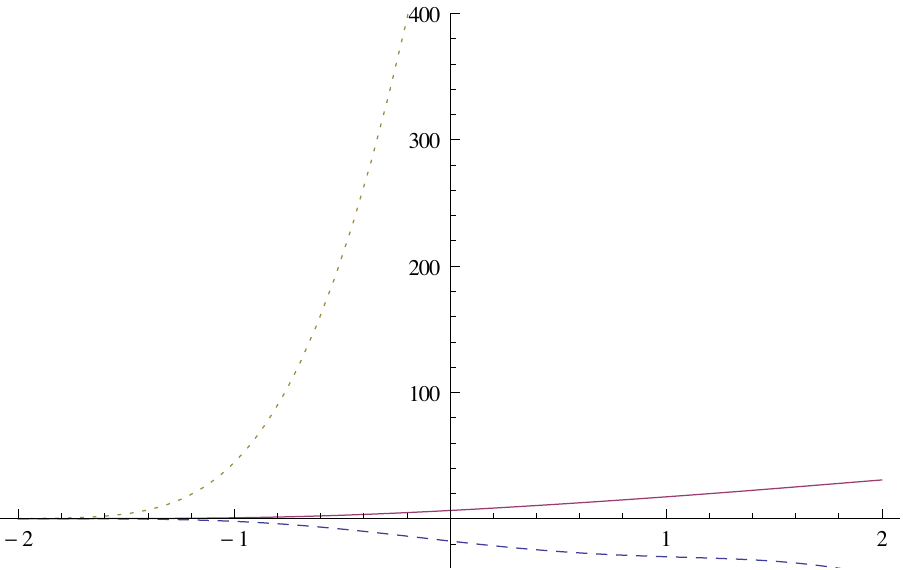}~~~
  \includegraphics[width= 0.45\textwidth]{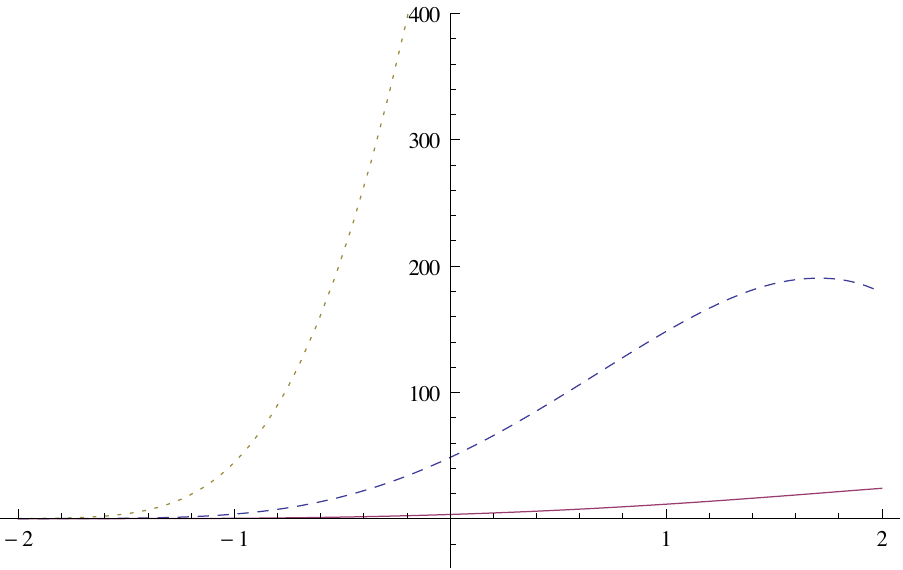} 

\caption{ The Bethe-Heitler(dotted), TCS(solid) and Interference(dashed) contributions to $\frac{d \sigma}{dQ^2dtdyd\phi}$ in $pb/\mbox{GeV}^4$ 
as a function of y (in CMS) for $Q^2 =4\mbox{GeV}^2$, $t =-0.1 \mbox{GeV}^2$, $\phi =0$ for the proton beam and $Pb$ target case.  }
\label{Fig:crosssection1}
\end{center}
\end{figure}

\begin{figure}[h]
\begin{center}
 \includegraphics[width= 0.45\textwidth]{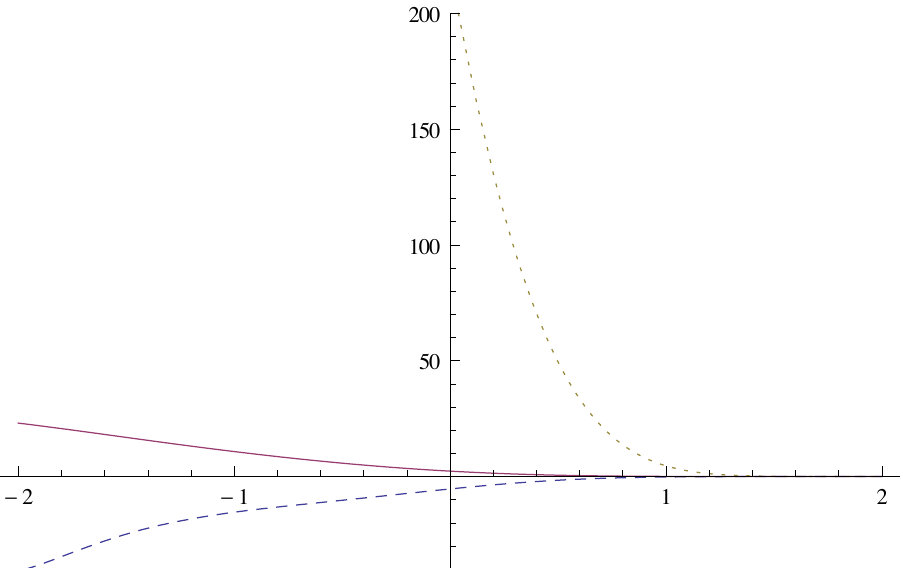} ~~~
 \includegraphics[width= 0.45\textwidth]{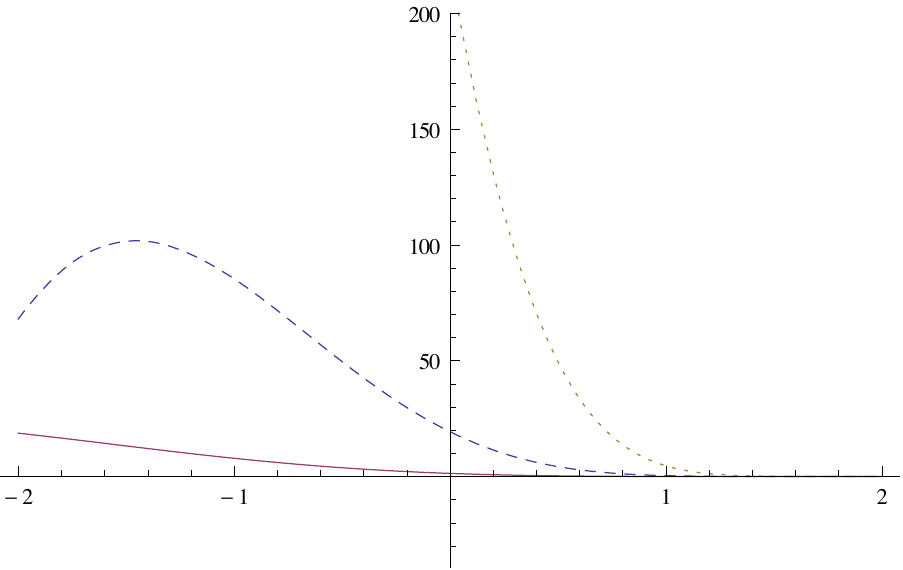} 
\caption{ The Bethe-Heitler(dotted), TCS(solid) and Interference(dashed) contributions to
 $\frac{d \sigma}{dQ^2dtdyd\phi}$ in $pb/\mbox{GeV}^4$ 
for BH(dotted), TCS(solid), Interference(dashed) 
as a function of y (in CMS) for $Q^2 =4\mbox{GeV}^2$, $t =-0.1 \mbox{GeV}^2$, $\phi =0$, for the $Pb$ beam and proton target case.}
\label{Fig:crosssection2}
\end{center}
\end{figure}

The main idea of the multi-purpose project AFTER@LHC is to extract the halo of the LHC proton or ion beams by means of  a bent-cristal and to use it as a beam in the fixed-target experiments \cite{AFTER}. The extracted beams will have a sufficiently high energy to produce on fixed target in ultraperipheral scattering a lepton pair with high invariant mass or heavy mesons. In these experiments a nucleus projectile or a nucleus target is treated as  a high-energy photon source which allows study of photon-hadron collisions. Fig. \ref{Fig:crosssection1} (and \ref{Fig:crosssection2} respectively) shows preliminary estimates of the Bethe-Heitler, TCS, and Interference contributions to the cross section, as functions of the CMS rapidity $y$, after integration over $\theta \in(\pi/4,3 \pi/4)$, in the region where the interference contribution is  best seen, for the collision of a proton beam with a $Pb$ target (and of a $Pb$ beam with a proton target respectively). The double distribution model of GPDs based on  MSTW08  parametrization is used.

\hspace*{2.cm} LO CF \hspace*{4cm} NLO CF
  
\hspace*{2.cm} LO CF \hspace*{4cm} NLO CF

\section{Summary and outlook
\label{sec4}}

We did not discuss here the rich phenomenology of DVCS and TCS processes which electron-ion colliders \cite{EIC} will allow to study. Neither did we comment on recent progresses in higher twist contributions \cite{HTwist} nor on the effect of  resummation of higher order QCD corrections  \cite{tolga}.  The physics of generalized parton distributions is definitely a domain of work in progress, both on the theory and on the experimental side.
\section*{Acknowledgments}
This work is  supported by  the Polish Grant NCN No DEC-
2011/01/D/ST2/03915, the French-Polish collaboration agreement Polonium, the ANR project "Partons", the COPIN-IN2P3 Agreement
 and the Joint
Research Activity "Study of Strongly Interacting Matter"
(HadronPhysics3, Grant Agreement n.283286) under the
7$^{th}$ Framework Programm of the European Community.



\bibliographystyle{apsrev4-1}


\end{document}